\documentclass[sigconf,nonacm]{acmart}

\usepackage{subfigure}
\usepackage{siunitx}
\usepackage{enumitem}

\usepackage{algorithm}
\usepackage[noend]{algpseudocode}
\let\ReturnInline\Return
\renewcommand{\Return}{\State\ReturnInline}
\algrenewcommand\algorithmicrequire{$\rhd$}
\algrenewcommand\algorithmicensure{$\square$}

\AtBeginDocument{%
  \providecommand\BibTeX{{%
    \normalfont B\kern-0.5em{\scshape i\kern-0.25em b}\kern-0.8em\TeX}}}

\setcopyright{acmcopyright}
\copyrightyear{2018}
\acmYear{2018}
\acmDOI{XXXXXXX.XXXXXXX}

\acmConference[Conference acronym 'XX]{Make sure to enter the correct
  conference title from your rights confirmation emai}{June 03--05,
  2018}{Woodstock, NY}

\newcommand{\ignore}[1]{}

\newcommand{\Lou}{\textit{Louvain}}

\begin{document}

\title[GVE-Louvain: Fast Louvain Algorithm for Community Detection in Shared Memory Setting]{GVE-Louvain: Fast Louvain Algorithm for\\Community Detection in Shared Memory Setting}


\author{Subhajit Sahu}
\email{subhajit.sahu@research.iiit.ac.in}
\affiliation{%
  \institution{IIIT Hyderabad}
  \streetaddress{Professor CR Rao Rd, Gachibowli}
  \city{Hyderabad}
  \state{Telangana}
  \country{India}
  \postcode{500032}
}


\settopmatter{printfolios=true}

\begin{abstract}
Community detection is the problem of identifying natural divisions in networks. Efficient parallel algorithms for identifying such divisions is critical in a number of applications, where the size of datasets have reached significant scales. This technical report presents one of the most efficient multicore implementations of the Louvain algorithm, a high quality community detection method. On a server equipped with dual 16-core Intel Xeon Gold 6226R processors, our Louvain, which we term as GVE-Louvain, outperforms Vite, Grappolo, NetworKit Louvain, and cuGraph Louvain (running on NVIDIA A100 GPU) by $50\times$, $22\times$, $20\times$, and $5.8\times$ faster respectively - achieving a processing rate of $560 M$ edges/s on a $3.8 B$ edge graph. In addition, GVE-Louvain improves performance at an average rate of $1.6\times$ for every doubling of threads.
\end{abstract}



\maketitle

\section{Introduction}
\label{sec:introduction}
Community detection\ignore{, also know as clustering,} is the problem of uncovering the underlying structure of complex networks, i.e., identifying groups of vertices that exhibit dense internal connections but sparse connections with the rest of the network, in an unsupervised manner. It is an NP-hard problem with numerous applications in domains such as drug discovery, protein annotation, topic discovery, anomaly detection, and criminal identification. Communities identified are intrinsic when based on network topology alone, and are disjoint when each vertex belongs to only one community \cite{com-gregory10}. One of the difficulties in the community detection problem is the lack of apriori knowledge on the number and size distribution of communities \cite{com-blondel08}. The \textit{Louvain} method \cite{com-blondel08} is a popular heuristic-based approach for community detection, with the modularity metric \cite{com-newman06} being used to measure the quality of communities identified.

In recent years, the collection of data and the relationships among them, represented as graphs, have reached unmatched levels. This has necessitated the design of efficient parallel algorithms for community detection on large networks. Existing studies on Louvain propose\ignore{a number of algorithmic} several optimizations \cite{com-rotta11, com-waltman13, com-gach14, com-traag15, com-lu15, com-ryu16, com-ozaki16, com-naim17, com-halappanavar17, com-ghosh18, com-traag19, com-zhang21, com-shi21, com-you22, com-aldabobi22} and parallelization techniques \cite{com-cheong13, com-wickramaarachchi14, com-lu15, com-zeng15, com-que15, com-naim17, com-fazlali17, com-halappanavar17, com-zeitz17, com-ghosh18, com-bhowmik19, com-gheibi20, com-shi21, com-bhowmick22}. Further, significant research effort has been focused on developing efficient parallel implementations of Louvain algorithm for multicore CPUs \cite{staudt2015engineering, staudt2016networkit, com-fazlali17, com-halappanavar17, qie2022isolate}, GPUs \cite{com-naim17}, CPU-GPU hybrids \cite{com-bhowmik19, com-mohammadi20}, multi-GPUs \cite{com-cheong13, kang2023cugraph, chou2022batched, com-gawande22}, and multi-node systems --- CPU only \cite{com-ghosh18, ghosh2018scalable, sattar2022scalable} / CPU-GPU hybrids \cite{com-bhowmick22}.

However, many of the aforementioned works concentrate on optimizing the local-moving phase of the Louvain algorithm, but do not address optimization for the aggregation phase of the algorithm, which emerges as a bottleneck after the local-moving phase has been optimized.\ignore{Some implementations also fail to adequately parallelize the algorithm.} These optimization techniques are also scattered over a number of papers, making it difficult for a reader to get a grip over them. Moreover, much attention has been directed towards GPU-based solutions. However, developing algorithms that efficiently utilize GPUs can be challenging both in terms of initial implementation and ongoing maintenance. Further, the soaring prices of GPUs present hurdles. The multicore/shared memory environment holds significance for community detection, owing to its energy efficiency and the prevalence of hardware with ample DRAM capacities. Through our implementation of the Louvain algorithm, we aim to underscore that CPUs remain adept at irregular computation\ignore{, especially for algorithms where workload diminishes progressively with each iteration}. Additionally, we show that achieving optimal performance necessitates a\ignore{heightened} focus on the\ignore{underlying} data structures\ignore{ rather than solely on algorithmic techniques, recognizing that algorithms are built upon these foundational structures}.

\ignore{Optimizing parallel community detection algorithms for modern hardware architectures can yield notable performance benefits and competitive advantages across applications. However, many of the current algorithms for community detection are challenging to parallelize due to their irregular and inherently sequential nature \cite{com-halappanavar17}, in addition to the complexities of handling concurrency, optimizing data access, reducing contention, minimizing load imbalance.}

In this\ignore{technical} report, we introduce our parallel multicore implementation of the Louvain algorithm\footnote{\url{https://github.com/puzzlef/louvain-communities-openmp}}. Our implementation employs asynchronous computation, utilizes parallel prefix sum and preallocated Compressed Sparse Row (CSR) data structures for identifying community vertices and storing the super-vertex graph during the aggregation phase, uses fast collision-free per-thread hash tables for the local-moving and aggregation phases, and incorporates an aggregation tolerance to avoid unnecessary aggregation phases. Additionally, we leverage established techniques such as threshold-scaling optimization, vertex pruning, limiting the number of iterations per pass, and OpenMP's \verb|dynamic| loop schedule. For each optimization, we determine optimal parameter settings, if there are any. To the best of our knowledge, our implementation represents the most efficient parallel Louvain algorithm implementation on multicore CPUs. We compare our implementation against other state-of-the-art implementations, including multi-core, multi-node, hybrid GPU, multi-GPU, and multi-node multi-GPU implementations, in Table \ref{tab:compare}. Both direct and indirect comparisons are provided, with details given in Sections \ref{sec:comparison} and \ref{sec:comparison-indirect} respectively.

\begin{table}[hbtp]
  \centering
  \caption{Speedup of our multicore implementation of Louvain algorithm compared to other state-of-the-art implementations. Direct comparisons are based on running the given implementation on our server, while indirect comparisons (denoted with a $*$, details given in Section \ref{sec:comparison-indirect}) involve comparing results obtained by the given implementation relative to a common reference (Grappolo). Among the Louvain implementations, some are multi-core, while others are categorized as multi-node (\textit{nodes}), hybrid GPU (\textit{GPU}), multi-GPU (\textit{GPUs, 1 GPU}), and multi-node multi-GPU (\textit{DGPUs}).}
  \label{tab:compare}
  \begin{tabular}{|c|c||c|}
    \toprule
    \textbf{Louvain implementation} &
    \textbf{Published} &
    \textbf{Our Speedup} \\
    \midrule
    Grappolo \cite{com-halappanavar17} & 2017 & $22\times$ \\ \hline
    Vite \cite{ghosh2018scalable} & 2018 & $50\times$ \\ \hline
    NetworKit Louvain \cite{staudt2016networkit} & 2016 & $20\times$ \\ \hline
    cuGraph Louvain \cite{kang2023cugraph} & 2023 & $5.8\times$ \\ \hline
    PLM \cite{staudt2015engineering} & 2015 & $48\times^*$ \\ \hline
    APLM \cite{com-fazlali17} & 2017 & $7.7\times^*$ \\ \hline
    Ghosh et al. (\ignore{8 }nodes) \cite{com-ghosh18} & 2018 & $5.9\times^*$ \\ \hline
    HyDetect (GPU) \cite{com-bhowmik19} & 2019 & $54\times^*$ \\ \hline
    Rundemanen (GPU) \cite{com-naim17} & 2017 & $9.8\times^*$ \\ \hline
    ACLM (GPU) \cite{com-mohammadi20} & 2020 & $6.0\times^*$ \\ \hline
    Nido (1 GPU) \cite{chou2022batched} & 2022 & $9.2\times^*$ \\ \hline
    cuVite (1 GPU) \cite{com-gawande22} & 2022 & $6.7\times^*$ \\ \hline
    Cheong et al. (\ignore{16/24 }GPUs) \cite{com-cheong13} & 2013 & $8.3\times^*$ \\ \hline
    Bhowmick et al. (\ignore{8 }DGPUs) \cite{com-bhowmick22} & 2022 & $4.9\times^*$ \\ \hline
  \bottomrule
  \end{tabular}
\end{table}

\section{Related work}
\label{sec:related}
The \textit{Louvain} method is a greedy modularity-optimization based community detection algorithm, and is introduced by Blondel et al. from the University of Louvain \cite{com-blondel08}. It identifies communities with resulting high modularity, and is thus widely favored \cite{com-lancichinetti09}. Algorithmic improvements proposed for the original algorithm include early pruning of non-promising candidates (leaf vertices) \cite{com-ryu16, com-halappanavar17, com-zhang21, com-you22}, attempting local move only on likely vertices \cite{com-ryu16, com-ozaki16, com-zhang21, com-shi21}, ordering of vertices based on node importance \cite{com-aldabobi22}, moving nodes to a random neighbor community \cite{com-traag15}, threshold scaling \cite{com-lu15, com-naim17, com-halappanavar17}, threshold cycling \cite{com-ghosh18}, subnetwork refinement \cite{com-waltman13, com-traag19}, multilevel refinement \cite{com-rotta11, com-gach14, com-shi21}, and early termination \cite{com-ghosh18}.

To parallelize the Louvain algorithm, a number of strategies have been attempted. These include using heuristics to break the sequential barrier \cite{com-lu15}, ordering vertices via graph coloring \cite{com-halappanavar17}, performing iterations asynchronously \cite{com-que15, com-shi21}, using adaptive parallel thread assignment \cite{com-fazlali17, com-naim17, com-sattar19, com-mohammadi20}, parallelizing the costly first iteration \cite{com-wickramaarachchi14}, using vector based hashtables \cite{com-halappanavar17}, and using sort-reduce instead of hashing \cite{com-cheong13}\ignore{, using simple partitions based of vertex ids \cite{com-cheong13, com-ghosh18}, and identifying and moving ghost/doubtful vertices \cite{com-zeng15, com-que15, com-bhowmik19, com-bhowmick22}}.\ignore{Platforms used range from an AMD multicore system \cite{com-fazlali17}, and Intel’s Knight's Landing, Haswell \cite{com-gheibi20}, SkylakeX, and Cascade Lake \cite{part-hossain21}. Other approaches include the use of MapReduce in a BigData batch processing framework \cite{com-zeitz17}.} It should however be noted though that community detection methods\ignore{such as the Louvain} that rely on modularity maximization are known to suffer from resolution limit problem \cite{com-ghosh19}\ignore{, which prevents identification of communities of certain sizes}.

We now discuss about a number of state-of-the-art implementation of Parallel Louvain. Mohammadi et al. \cite{com-mohammadi20} propose the Adaptive CUDA Louvain Method (ACLM), employing GPU acceleration. Their approach involves evaluating the change in modularity for each edge in the graph, reflecting the impact of moving the source vertex to the community of the target vertex. Notably, only the computation of modularity changes occurs on the GPU, with other algorithmic steps executed on the CPU. Sattar and Arifuzzaman \cite{sattar2022scalable} discuss their Distributed Parallel Louvain Algorithm with Load-balancing (DPLAL). They begin by partitioning the input graph using METIS, employing the edge-cut partitioning approach. In each iteration, they determine the optimal community to relocate to for each vertex, execute community changes while tackling duality concerns, evaluate the new modularity, and generate the subsequent level graph. This iterative process continues until no further enhancement in modularity is achieved. Nonetheless, they conduct only a single iteration of the local-moving phase per iteration, which might result in the identification of substandard quality communities.\ignore{It is not clear if they perform local-moves to the best community per vertex, or just any community with a positive delta-modularity score. They do not present the quality of returned communities.} Ghosh et al. \cite{ghosh2018scalable} propose Vite, a distributed memory parallel implementation of the Louvain method that incorporates several heuristics to enhance performance while maintaining solution quality, while Grappolo, by Halappanavar et al. \cite{com-halappanavar17}, is a shared memory parallel implementation. Qie et al. \cite{qie2022isolate} present a graph partitioning algorithm that divides the graph into sets of partitions, aiming to minimize inter-partition communication delay and avoid community swaps, akin to the graph coloring approach proposed by Halappanavar et al. \cite{com-halappanavar17}. We do not observe the community swap issue on multicore CPUs (it likely resolves itself), but do observe it on GPUs (likely due to lockstep execution\ignore{of the steps of the algorithm}).

Bhowmick et al. \cite{com-bhowmik19} introduce HyDetect, a community detection algorithm designed for hybrid CPU-GPU systems, employing a divide-and-conquer strategy. The algorithm partitions the graph among the CPU and GPU components of a node, facilitating independent community detection utilizing the Louvain algorithm in both segments. Bhowmick et al. \cite{com-bhowmick22} later introduce a multi-node multi-GPU Louvain community detection algorithm, which involves graph partitioning across multiple nodes, refinement through identification and migration of doubtful vertices between processors, and utilization of a hierarchical merging algorithm ensuring that the merged components can be accommodated within a processor at any given point. Chou and Ghosh \cite{chou2022batched} present Nido, a batched clustering method for GPUs that leverages a partition-based design, and can process graphs larger than the combined GPU memory of a node. Gawande et al. \cite{com-gawande22} propose cuVite, an ongoing endeavor focusing on distributed Louvain for heterogeneous systems --- building upon their previous work involving parallelizing\ignore{the} Louvain\ignore{method} for community detection on CPU-only distributed systems. NetworKit \cite{staudt2016networkit} is a software package designed for analyzing the structural aspects of graph data sets with billions of connections. It is implemented as a hybrid with C++ kernels and a Python frontend, and includes a parallel implementation of the Louvain algorithm. Finally, cuGraph \cite{kang2023cugraph} is a GPU-accelerated graph analytics library that is part of the RAPIDS suite of data science and machine learning tools. It harnesses the power of NVIDIA GPUs to significantly speed up graph analytics compared to traditional CPU-based methods. cuGraph's core is written in C++ with CUDA and is primarily accessed through a Python interface\ignore{, making it user-friendly for data scientists and developers working in Python}.

\ignore{A few open source implementations and software packages have been developed for community detection. Vite \cite{ghosh2018scalable} is a distributed memory parallel implementation of the Louvain method that incorporates several heuristics to enhance performance while maintaining solution quality, while Grappolo \cite{com-halappanavar17} is a shared memory parallel implementation. NetworKit \cite{staudt2016networkit} is a software package designed for analyzing the structural aspects of graph data sets with billions of connections. It is implemented as a hybrid with C++ kernels and a Python frontend, and includes a parallel implementation of the Louvain algorithm.}

However, most existing works only focus on optimizing the local-moving phase of the Louvain algorithm, and lack effective parallelization. For instance, the implementation of NetworKit Louvain exhibits several shortcomings. It employs plain OpenMP parallelization for certain operations, utilizing a static schedule with a chunk size of 1, which may not be optimal when threads are writing to adjacent memory addresses. Additionally, NetworKit Louvain employs guided scheduling for the local-moving phase, whereas we utilize dynamic scheduling for better performance. NetworKit Louvain also generates a new graph for each coarsening step, leading to repeated memory allocation and preprocessing due to recursive calls. The coarsening\ignore{process} involves several sequential operations, and adding each edge to the coarsened graph requires $O(D)$ operations, where $D$ represents the average degree of a vertex, which is suboptimal for parallelism. Lastly, NetworKit Louvain lacks parallelization for flattening the dendrogram, potentially hindering its performance\ignore{in certain scenarios}.\ignore{Moreover, NetworKit Louvain only times the local-moving and aggregation phases, underestimating its performance compared to our implementation.} Our parallel implementation of Louvain addresses these issues.

\section{Preliminaries}
\label{sec:preliminaries}
Consider an undirected graph $G(V, E, w)$ with $V$ representing the set of vertices, $E$ the set of edges, and $w_{ij} = w_{ji}$ denoting the weight associated with each edge. In the case of an unweighted graph, we assume unit weight for each edge ($w_{ij} = 1$). Additionally, the neighbors of a vertex $i$ are denoted as $J_i = \{j\ |\ (i, j) \in E\}$, the weighted degree of each vertex as $K_i = \sum_{j \in J_i} w_{ij}$, the total number of vertices as $N = |V|$, the total number of edges as $M = |E|$, and the sum of edge weights in the undirected graph as $m = \sum_{i, j \in V} w_{ij}/2$.

\subsection{Community detection}

Disjoint community detection is the process of identifying a community membership mapping, $C: V \rightarrow \Gamma$, where each vertex $i \in V$ is assigned a community-id $c \in \Gamma$, where $\Gamma$ is the set of community-ids. We denote the vertices of a community $c \in \Gamma$ as $V_c$, and the community that a vertex $i$ belongs to as $C_i$. Further, we denote the neighbors of vertex $i$ belonging to a community $c$ as $J_{i \rightarrow c} = \{j\ |\ j \in J_i\ and\ C_j = c\}$, the sum of those edge weights as $K_{i \rightarrow c} = \sum_{j \in J_{i \rightarrow c}} w_{ij}$, the sum of weights of edges within a community $c$ as $\sigma_c = \sum_{(i, j) \in E\ and\ C_i = C_j = c} w_{ij}$, and the total edge weight of a community $c$ as $\Sigma_c = \sum_{(i, j) \in E\ and\ C_i = c} w_{ij}$ \cite{com-leskovec21}.

\subsection{Modularity}

Modularity serves as a\ignore{fitness} metric for assessing the quality of communities identified by heuristic-based community detection algorithms. It is computed as the difference between the fraction of edges within communities and the expected fraction if edges were randomly distributed, yielding a range of $[-0.5, 1]$ where higher values indicate better results \cite{com-brandes07}.\ignore{The optimization of this metric theoretically leads to the optimal grouping \cite{com-newman04, com-traag11}.} The modularity $Q$ of identified communities is determined using Equation \ref{eq:modularity}, where $\delta$ represents the Kronecker delta function ($\delta (x,y)=1$ if $x=y$, $0$ otherwise). The \textit{delta modularity} of moving a vertex $i$ from community $d$ to community $c$, denoted as $\Delta Q_{i: d \rightarrow c}$, can be computed using Equation \ref{eq:delta-modularity}.

\begin{equation}
\label{eq:modularity}
  Q
  = \frac{1}{2m} \sum_{(i, j) \in E} \left[w_{ij} - \frac{K_i K_j}{2m}\right] \delta(C_i, C_j)
  = \sum_{c \in \Gamma} \left[\frac{\sigma_c}{2m} - \left(\frac{\Sigma_c}{2m}\right)^2\right]
\end{equation}

\begin{equation}
\label{eq:delta-modularity}
  \Delta Q_{i: d \rightarrow c}
  = \frac{1}{m} (K_{i \rightarrow c} - K_{i \rightarrow d}) - \frac{K_i}{2m^2} (K_i + \Sigma_c - \Sigma_d)
\end{equation}

\subsection{Louvain algorithm}
\label{sec:about-louvain}

The Louvain method \cite{com-blondel08} is a modularity optimization based agglomerative algorithm for identifying high quality disjoint communities in large networks. It has a time complexity of $O (L |E|)$ (with $L$ being the total number of iterations performed), and a space complexity of $O(|V| + |E|)$ \cite{com-lancichinetti09}. The algorithm consists of two phases: the \textit{local-moving phase}, where each vertex $i$ greedily decides to move to the community of one of its neighbors $j \in J_i$ that gives the greatest increase in modularity $\Delta Q_{i:C_i \rightarrow C_j}$ (using Equation \ref{eq:delta-modularity}), and the \textit{aggregation phase}, where all the vertices in a community are collapsed into a single super-vertex. These two phases make up one pass, which repeats until there is no further increase in modularity \cite{com-blondel08, com-leskovec21}. We observe that Louvain obtains high-quality communities, with $3.0 - 30\%$ higher modularity than that obtained by LPA, but requires $2.3 - 14\times$ longer to converge.

\section{Approach}
\label{sec:approach}
\subsection{Optimizations for Louvain algorithm}
\label{sec:louvain}

We use a parallel implementation of the Louvain method to determine suitable parameter settings and optimize the original algorithm through experimentation with a variety of techniques. We use \textit{asynchronous} version of Louvain, where threads work independently on different parts of the graph. This allows for faster convergence but can also lead to more variability in the final result \cite{com-blondel08, com-halappanavar17}. Further, we allocate a separate hashtable per thread to keep track of the delta-modularity of moving to each community linked to a vertex in the local-moving phase of the algorithm, and to keep track of the total edge weight from one super-vertex to the other super-vertices in the aggregation phase of the algorithm.

Our optimizations include using OpenMP's \verb|dynamic| loop schedule, limiting the number of iterations per pass to $20$, using a tolerance drop rate of $10$, setting an initial tolerance of $0.01$, using an aggregation tolerance of $0.8$, employing vertex pruning, making use of parallel prefix sum and preallocated Compressed Sparse Row (CSR) data structures for finding community vertices and for storing the super-vertex graph during the aggregation phase, and using fast collision-free per-thread hashtables which are well separated in their memory addresses (\textit{Far-KV}) for the local-moving and aggregation phases of the algorithm. Details\ignore{ on each of the optimizations is} are given below.

For each optimization, we test a number of relevant alternatives, and show the relative time and the relative modularity of communities obtained by each alternative in Figure \ref{fig:louvain-opt}. This result is obtained by running the tests on each graph in the dataset (see Table \ref{tab:dataset}), $5$ times on each graph to reduce the impact of noise, taking their geometric mean and arithmetic mean for the runtime and modularity respectively, and representing them as a ratio within each optimization category.

\subsubsection{Adjusting OpenMP loop schedule}

We attempt \textit{static}, \textit{dynamic}, \textit{guided}, and \textit{auto} loop scheduling approaches of OpenMP (each with a chunk size of $2048$) to parallelize the local-moving and aggregation phases of the Louvain algorithm. Results indicate that the scheduling behavior can have small impact on the quality of obtained communities. We consider OpenMP's \verb|dynamic| loop schedule to be the best choice, as it helps achieve better load balancing when the degree distribution of vertices is non-uniform, and offers a $7\%$ reduction in runtime with respect to OpenMP's \textit{auto} loop schedule, with only a $0.4\%$ reduction in the modularity of communities obtained (which is likely to be just noise).

\subsubsection{Limiting the number of iterations per pass}

Restricting the number of iterations of the local-moving phase ensures its termination within a reasonable number of iterations, which helps minimize runtime. This can be important since the local-moving phase performed in the first pass is the most expensive step of the algorithm. However, choosing too small a limit may worsen convergence rate. Our results indicate that limiting the maximum number of iterations to $20$ allows for $13\%$ faster convergence, when compared to a maximum iterations of $100$.

\subsubsection{Adjusting tolerance drop rate (threshold scaling)}

Tolerance is used to detect convergence in the local-moving phase of the Louvain algorithm, i.e., when the total delta-modularity in a given iteration is below or equal to the specified tolerance, the local-moving phase is considered to have converged. Instead of using a fixed tolerance across all passes of the Louvain algorithm, we can start with an initial high tolerance and then gradually reduce it. This is known as threshold scaling \cite{com-lu15, com-naim17, com-halappanavar17}, and it helps minimize runtime of the first pass of the algorithm (which is usually the most expensive). Based on our findings, a tolerance drop rate of $10$ yields $4\%$ faster convergence, with respect to a tolerance drop rate of $1$ (threshold scaling disabled), with no reduction in quality of communities obtained.

\subsubsection{Adjusting initial tolerance}

Starting with a smaller initial tolerance allows the algorithm to explore broader possibilities for community assignments in the early stage, but comes at the cost of increased runtime. We find an initial tolerance of $0.01$ provides a $14\%$ reduction in runtime of the algorithm with no reduction in the quality of identified communities, when compared to an initial tolerance of $10^{-6}$.

\subsubsection{Adjusting aggregation tolerance}

The aggregation tolerance determines the point at which communities are considered to have converged based on the number of community merges. In other words, if too few communities merged this pass we should stop here, i.e., if $|V_{aggregated}|/|V| \geq$ aggregation tolerance, we consider the algorithm to have converged. Adjusting aggregation tolerance allows the algorithm to stop earlier when further merges have minimal impact on the final result. According to our observations, an aggregation tolerance of $0.8$ appears to be the best choice, as it presents a $14\%$ reduction in runtime, when compared to the aggregation tolerance being disabled ($1$), while identifying final communities of equivalent quality.

\subsubsection{Vertex pruning}

\begin{figure}[hbtp]
  \centering
  \subfigure{
    \label{fig:louvain-pruning--all}
    \includegraphics[width=0.78\linewidth]{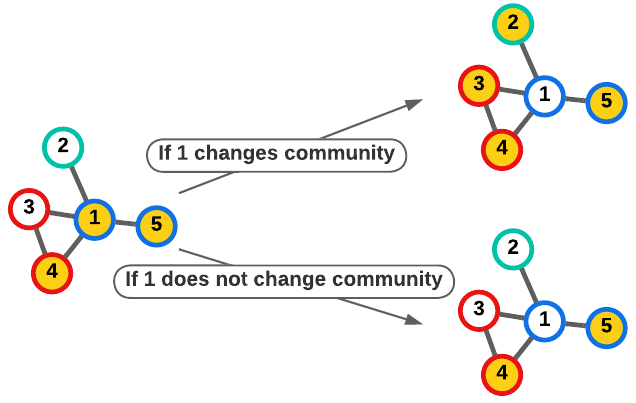}
  } \\[-2ex]
  \caption{Illustration of vertex pruning optimization: After processing vertex $1$, it's unmarked. If vertex $1$ changes its community, its neighbors are marked for processing. Community membership of each vertex is depicted by border color, and marked vertices are highlighted in yellow.}
  \label{fig:louvain-pruning}
\end{figure}

Vertex pruning is a technique that is used to minimize unnecessary computation \cite{com-ryu16, com-ozaki16, com-zhang21, com-shi21}. Here, when a vertex changes its community, its marks its neighbors to be processed. Once a vertex has been processed, it is marked as not to be processed. However, it comes with an added overhead of marking an unmarking of vertices. Based on our results, vertex pruning justifies this overhead, and should be enabled for $11\%$ improvement in performance. An illustration of this\ignore{vertex pruning optimization} is shown in Figure \ref{fig:louvain-pruning}.

\begin{figure}[hbtp]
  \centering
  \subfigure{
    \label{fig:louvain-hashtable--all}
    \includegraphics[width=0.88\linewidth]{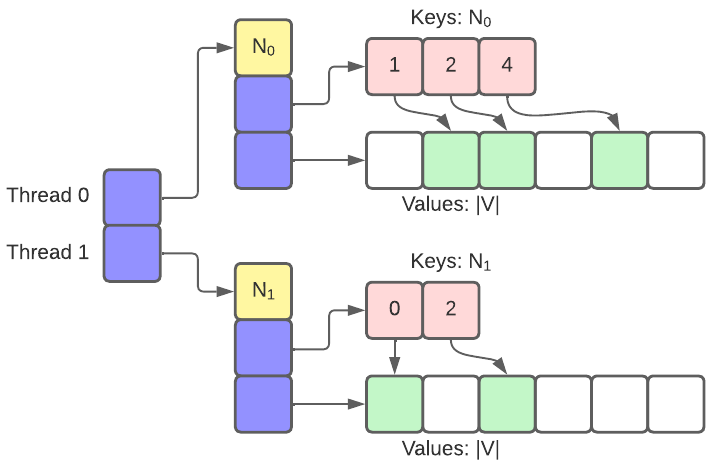}
  } \\[-2ex]
  \caption{Illustration of collision-free per-thread hashtables that are well separated in their memory addresses (Far-KV), for two threads. Each hashtable consists of a keys vector, values vector (of size $|V|$), and a key count ($N_0$/$N_1$). The value associated with each key is stored/accumulated in the index pointed by the key. As the key count of each hashtable is updated independently, we allocate it separately on the heap to avoid false cache sharing. These are used in the local-moving and aggregation phases of our Louvain implementation.}
  \label{fig:louvain-hashtable}
\end{figure}

\begin{figure*}[hbtp]
  \centering
  \subfigure{
    \label{fig:louvain-opt--all}
    \includegraphics[width=0.98\linewidth]{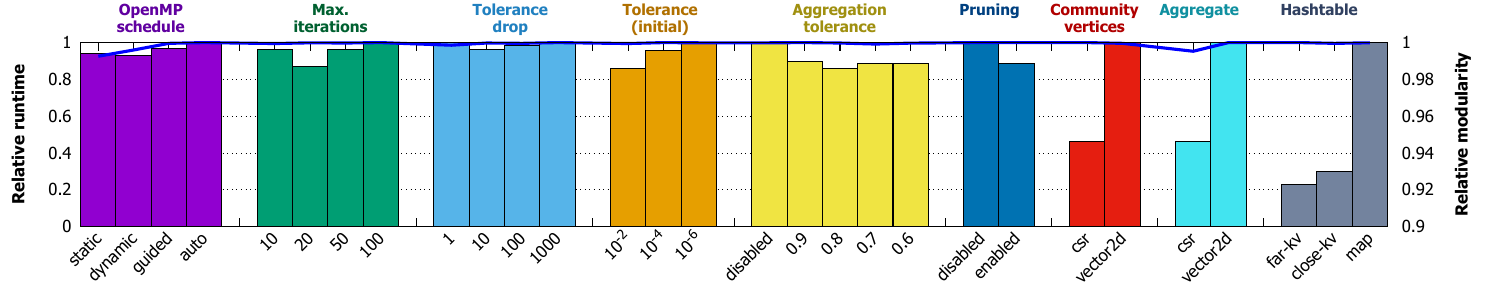}
  } \\[-2ex]
  \caption{Impact of various parameter controls and optimizations on the runtime and result quality (modularity) of the \Lou{} algorithm. We show the impact of each optimization upon the relative runtime on the left Y-axis, and upon the relative modularity on the right Y-axis.}
  \label{fig:louvain-opt}
\end{figure*}

\subsubsection{Finding community vertices for aggregation phase}

In the aggregation phase of the Louvain algorithm, the communities obtained in the previous local-moving phase of the algorithm are combined into super-vertices in the aggregated graph, with the edges between two super-vertices being equal to the total weight of edges between the respective communities. This requires one to obtain the list of vertices belonging to each community, instead of the mapping of community membership of each vertex that we have after the local-moving phase ends. A straight-forward implementation of this would make use of two-dimensional arrays for storing vertices belonging to each community, with the index in the first dimension representing the community id $c$, and the index in the second dimension pointing to the $n^{th}$ vertex in the given community $c$. However, this requires memory allocation during the algorithm, which is expensive. Employing a parallel prefix sum technique along with a preallocated Compressed Sparse Row (CSR) data structure eliminates repeated memory allocation and deallocation, enhancing performance. Indeed, out findings indicate that using parallel prefix sum along with a preallocated CSR is $2.2\times$ faster than using 2D arrays for aggregating vertices.

\subsubsection{Storing aggregated communities (super-vertex graph)}

After the list of vertices belonging to each community have been obtained, the communities need to be aggregated (or compressed) into super-vertices, such that edges between two super-vertices being equal to the total weight of edges between the respective communities. This is generally called the super-vertex graph, or the compressed graph. It is then used as an input to the local-moving phase of the next pass of the Louvain algorithm. A simple data structure to store the super-vertex graph in the adjacency list format would be a two-dimensional array. Again, this requires memory allocation during the algorithm, which is bad for performance. Utilizing two preallocated CSRs, one for the source graph and the other for the target graph (except the first pass, where the dynamic graph may be stored in any desired format suitable for dynamic batch updates), along with parallel prefix sum can help here. We observe that using parallel prefix sum along with preallocated CSRs for maintaining the super-vertex graph is again $2.2\times$ faster than using 2D arrays.

\subsubsection{Hashtable design for local-moving/aggregation phases}

One can use C++'s inbuilt maps as per-thread (independent) hashtables for the Louvain algorithm. But this has poor performance. So we use a key-list and a full-size values array (collision-free) to dramatically improve performance. However, if the memory addresses of the hashtables are nearby (\textit{Close-KV}) --- as with NetworKit Louvain \cite{staudt2016networkit}, even if each thread uses its own hashtable exclusively, performance is not as high. This is possibly due to false cache-sharing. Alternatively, if we ensure that the memory address of each hashtable are farther away (\textit{Far-KV}), the performance improves. Our results indicate that \textit{Far-KV} has the best performance and is $4.4\times$ faster than \textit{Map}, and $1.3\times$ faster than \textit{Close-KV} hashtable implementations. An illustration of \textit{Far-KV} hashtable is shown in Figure \ref{fig:louvain-hashtable}.

\subsection{Our optimized Louvain implementation}

\begin{figure*}[hbtp]
  \centering
  \subfigure{
    \label{fig:louvain-pass--all}
    \includegraphics[width=0.98\linewidth]{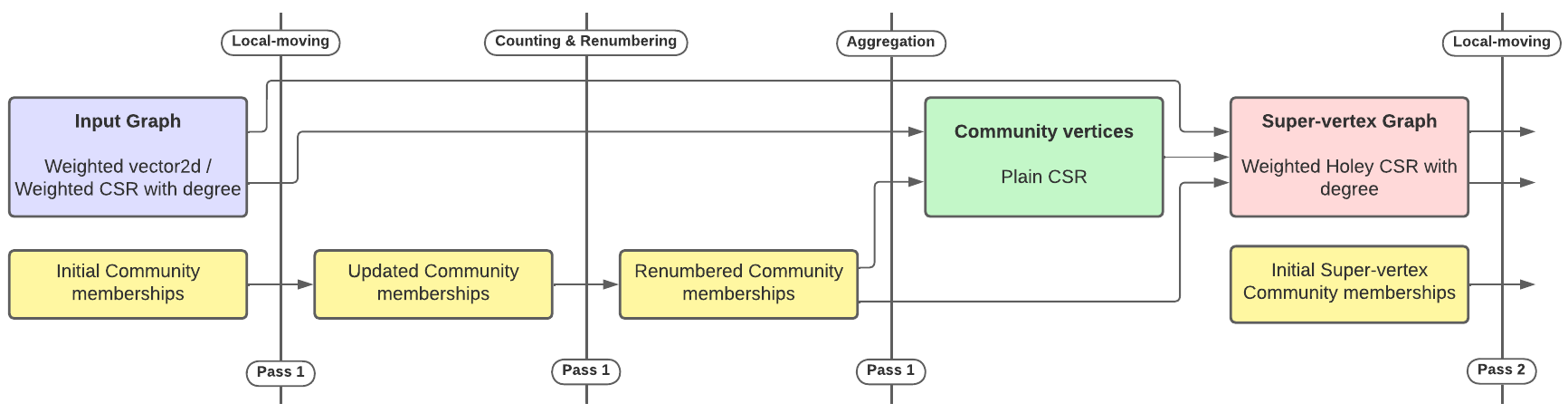}
  } \\[-2ex]
  \caption{A flow diagram illustrating the first pass of GVE-Louvain for a Weighted 2D-vector based or a Weighted CSR with degree based input graph. In the local-moving phase, vertex community memberships are updated until the total change in delta-modularity across all vertices reaches a specified threshold. Community memberships are then counted and renumbered. In the aggregation phase, community vertices in a CSR are first obtained. This is used to create the super-vertex graph stored in a Weighted Holey CSR with degree. In subsequent passes, the input is a Weighted Holey CSR with degree and initial community membership for super-vertices from the previous pass.}
  \label{fig:louvain-pass}
\end{figure*}

We now explain the implementation of GVE-Louvain in Algorithms \ref{alg:louvain}, \ref{alg:louvainlm}, and \ref{alg:louvainag}. We aim to incorporate GVE-Louvain into our upcoming command-line graph processing tool named "GVE", derived from "Graph(Vertices, Edges)". GVE-Louvain exhibits a time complexity of $O(KM)$, where $K$ signifies the total iterations conducted. Its space complexity is $O(TN + M)$, where $T$ denotes the number of threads utilized, and $TN$ represents the collision-free hash tables $H_t$ employed per thread. A flow diagram illustrating the first pass of GVE-Louvain is shown in Figure \ref{fig:louvain-pass}.

\subsubsection{Main step of GVE-Louvain}

The main step of GVE-Louvain (\texttt{louvian()} function) is outlined in Algorithm \ref{alg:louvain}. It encompasses initialization, the local-moving phase, and the aggregation phase. Here, the \texttt{louvain()} function takes the input graph $G$, and returns the community membership $C$ for each vertex. In line \ref{alg:louvain--initialization}, we first initialize the community membership $C$ of each vertex in $G$, and perform passes of the Louvain algorithm, limited to $MAX\_PASSES$ (lines \ref{alg:louvain--passes-begin}-\ref{alg:louvain--passes-end}). In each pass, we initialize the total edge weight of each vertex $K'$, the total edge weight of each community $\Sigma'$, and the community membership $C'$ of each vertex in the current graph $G'$ (line \ref{alg:louvain--reset-weights}); and mark all vertices as unprocessed (line \ref{alg:louvain--reset-affected}).

\begin{algorithm}[hbtp]
\caption{GVE-Louvain: Our parallel Louvain algorithm.}
\label{alg:louvain}
\begin{algorithmic}[1]
\Require{$G$: Input graph}
\Require{$C$: Community membership of each vertex}
\Require{$G'$: Input/super-vertex graph}
\Require{$C'$: Community membership of each vertex in $G'$}
\Require{$K'$: Total edge weight of each vertex}
\Require{$\Sigma'$: Total edge weight of each community}
\Ensure{$H_t$: Collision-free per-thread hashtable}
\Ensure{$l_i$: Number of iterations performed (per pass)}
\Ensure{$l_p$: Number of passes performed}
\Ensure{$\tau$: Per iteration tolerance}
\Ensure{$\tau_{agg}$: Aggregation tolerance}

\Statex

\Function{louvain}{$G$} \label{alg:louvain--begin}
  \State Vertex membership: $C \gets [0 .. |V|)$ \textbf{;} $G' \gets G$ \label{alg:louvain--initialization}
  \ForAll{$l_p \in [0 .. \text{\small{MAX\_PASSES}})$} \label{alg:louvain--passes-begin}
    \State $\Sigma' \gets K' \gets vertexWeights(G')$ \textbf{;} $C' \gets [0 .. |V'|)$ \label{alg:louvain--reset-weights}
    \State Mark all vertices in $G'$ as unprocessed \label{alg:louvain--reset-affected}
    \State $l_i \gets louvainMove(G', C', K', \Sigma')$ \Comment{Alg. \ref{alg:louvainlm}} \label{alg:louvain--local-move}
    \If{$l_i \le 1$} \textbf{break} \Comment{Globally converged?} \label{alg:louvain--globally-converged}
    \EndIf
    \State $|\Gamma|, |\Gamma_{old}| \gets$ Number of communities in $C$, $C'$
    \If{$|\Gamma|/|\Gamma_{old}| > \tau_{agg}$} \textbf{break} \Comment{Low shrink?} \label{alg:louvain--aggregation-tolerance}
    \EndIf
    \State $C' \gets$ Renumber communities in $C'$ \label{alg:louvain--renumber}
    \State $C \gets$ Lookup dendrogram using $C$ to $C'$ \label{alg:louvain--lookup}
    \State $G' \gets louvainAggregate(G', C')$ \Comment{Alg. \ref{alg:louvainag}} \label{alg:louvain--aggregate}
    \State $\tau \gets \tau / \text{\small{TOLERANCE\_DROP}}$ \Comment{Threshold scaling} \label{alg:louvain--threshold-scaling}
  \EndFor \label{alg:louvain--passes-end}
  \State $C \gets$ Lookup dendrogram using $C$ to $C'$ \label{alg:louvain--lookup-last}
  \Return{$C$} \label{alg:louvain--return}
\EndFunction \label{alg:louvain--end}
\end{algorithmic}
\end{algorithm}

\begin{algorithm}[hbtp]
\caption{Local-moving phase of GVE-Louvain.}
\label{alg:louvainlm}
\begin{algorithmic}[1]
\Require{$G'$: Input/super-vertex graph}
\Require{$C'$: Community membership of each vertex}
\Require{$K'$: Total edge weight of each vertex}
\Require{$\Sigma'$: Total edge weight of each community}
\Ensure{$H_t$: Collision-free per-thread hashtable}
\Ensure{$l_i$: Number of iterations performed}
\Ensure{$\tau$: Per iteration tolerance}

\Statex

\Function{louvainMove}{$G', C', K', \Sigma'$} \label{alg:louvainlm--move-begin}
  \ForAll{$l_i \in [0 .. \text{\small{MAX\_ITERATIONS}})$} \label{alg:louvainlm--iterations-begin}
    \State Total delta-modularity per iteration: $\Delta Q \gets 0$ \label{alg:louvainlm--init-deltaq}
    \ForAll{unprocessed $i \in V'$ \textbf{in parallel}} \label{alg:louvainlm--loop-vertices-begin}
      \State Mark $i$ as processed (prune) \label{alg:louvainlm--prune}
      \State $H_t \gets scanCommunities(\{\}, G', C', i, false)$ \label{alg:louvainlm--scan}
      \State $\rhd$ Use $H_t, K', \Sigma'$ to choose best community
      \State $c^* \gets$ Best community linked to $i$ in $G'$ \label{alg:louvainlm--best-community-begin}
      \State $\delta Q^* \gets$ Delta-modularity of moving $i$ to $c^*$ \label{alg:louvainlm--best-community-end}
      \If{$c^* = C'[i]$} \textbf{continue} \label{alg:louvainlm--best-community-same}
      \EndIf
      \State $\Sigma'[C'[i]] -= K'[i]$ \textbf{;} $\Sigma'[c^*] += K'[i]$ \textbf{atomic} \label{alg:louvainlm--perform-move-begin}
      \State $C'[i] \gets c^*$ \textbf{;} $\Delta Q \gets \Delta Q + \delta Q^*$ \label{alg:louvainlm--perform-move-end}
      \State Mark neighbors of $i$ as unprocessed \label{alg:louvainlm--remark}
    \EndFor \label{alg:louvainlm--loop-vertices-end}
    \If{$\Delta Q \le \tau$} \textbf{break} \Comment{Locally converged?} \label{alg:louvainlm--locally-converged}
    \EndIf
  \EndFor \label{alg:louvainlm--iterations-end}
  \Return{$l_i$} \label{alg:louvainlm--return}
\EndFunction \label{alg:louvainlm--move-end}

\Statex

\Function{scanCommunities}{$H_t, G', C', i, self$}
  \ForAll{$(j, w) \in G'.edges(i)$}
    \If{\textbf{not} $self$ and $i = j$} \textbf{continue}
    \EndIf
    \State $H_t[C'[j]] \gets H_t[C'[j]] + w$
  \EndFor
  \Return{$H_t$}
\EndFunction
\end{algorithmic}
\end{algorithm}

\begin{algorithm}[hbtp]
\caption{Aggregation phase of GVE-Louvain.}
\label{alg:louvainag}
\begin{algorithmic}[1]
\Require{$G'$: Input/super-vertex graph}
\Require{$C'$: Community membership of each vertex}
\Ensure{$G'_{C'}$: Community vertices (CSR)}
\Ensure{$G''$: Super-vertex graph (weighted CSR)}
\Ensure{$*.offsets$: Offsets array of a CSR graph}
\Ensure{$H_t$: Collision-free per-thread hashtable}

\Statex

\Function{louvainAggregate}{$G', C'$}
  \State $\rhd$ Obtain vertices belonging to each community
  \State $G'_{C'}.offsets \gets countCommunityVertices(G', C')$ \label{alg:louvainag--coff-begin}
  \State $G'_{C'}.offsets \gets exclusiveScan(G'_{C'}.offsets)$ \label{alg:louvainag--coff-end}
  \ForAll{$i \in V'$ \textbf{in parallel}} \label{alg:louvainag--comv-begin}
    \State Add edge $(C'[i], i)$ to CSR $G'_{C'}$ atomically
  \EndFor \label{alg:louvainag--comv-end}
  \State $\rhd$ Obtain super-vertex graph
  \State $G''.offsets \gets communityTotalDegree(G', C')$ \label{alg:louvainag--yoff-begin}
  \State $G''.offsets \gets exclusiveScan(G''.offsets)$ \label{alg:louvainag--yoff-end}
  \State $|\Gamma| \gets$ Number of communities in $C'$
  \ForAll{$c \in [0, |\Gamma|)$ \textbf{in parallel}} \label{alg:louvainag--y-begin}
    \If{degree of $c$ in $G'_{C'} = 0$} \textbf{continue}
    \EndIf
    \State $H_t \gets \{\}$
    \ForAll{$i \in G'_{C'}.edges(c)$}
      \State $H_t \gets scanCommunities(H, G', C', i, true)$
    \EndFor
    \ForAll{$(d, w) \in H_t$}
      \State Add edge $(c, d, w)$ to CSR $G''$ atomically
    \EndFor
  \EndFor \label{alg:louvainag--y-end}
  \Return $G''$ \label{alg:louvainag--return}
\EndFunction
\end{algorithmic}
\end{algorithm}

Next, in line \ref{alg:louvain--local-move}, we perform the local-moving phase by calling \texttt{louvainMove()} (Algorithm $\ref{alg:louvainlm}$), which optimizes community assignments. If the local-moving phase converged in a single iteration, global convergence is implied and we terminate the passes (line \ref{alg:louvain--globally-converged}). Further, if the drop is community count $|\Gamma|$ is too small, we have reached the point of diminishing returns, and thus stop at the current pass (line \ref{alg:louvain--aggregation-tolerance}).

In case convergence has not been achieved, we renumber communities (line \ref{alg:louvain--renumber}), update top-level community memberships $C$ with dendrogram lookup (line \ref{alg:louvain--lookup}), perform the aggregation phase by calling \texttt{louvainAggregate()} (Algorithm \ref{alg:louvainag}), and scale the convergence threshold for subsequent passes, i.e., perform threshold scaling (line \ref{alg:louvain--threshold-scaling}). The next pass continues in line \ref{alg:louvain--passes-begin}. At the end of all passes, we perform a final update of the top-level community memberships $C$ with dendrogram lookup (line \ref{alg:louvain--lookup-last}), and return the top-level community membership $C$ of each vertex in $G$.

\subsubsection{Local-moving phase of GVE-Louvain}

The pseuodocode for the local-moving phase of GVE-Louvain is presented in Algorithm \ref{alg:louvainlm}, which iteratively moves vertices between communities to maximize modularity. Here, the \texttt{louvainMove()} function takes the current graph $G'$, community membership $C'$, total edge weight of each vertex $K'$, and total edge weight of each community $\Sigma'$ as input, and returns the number of iterations performed $l_i$.

Lines \ref{alg:louvainlm--iterations-begin}-\ref{alg:louvainlm--iterations-end} represent the main loop of the local-moving phase. In line \ref{alg:louvainlm--init-deltaq}, we first initialize the total delta-modularity per iteration $\Delta Q$. Next, in lines \ref{alg:louvainlm--loop-vertices-end}-\ref{alg:louvainlm--loop-vertices-end}, we iterate over unprocessed vertices in parallel. For each vertex $i$, we mark $i$ as processed - vertex pruning (line \ref{alg:louvainlm--prune}), scan communities connected to $i$ - excluding self (line \ref{alg:louvainlm--scan}), determine the best community $c*$ to move $i$ to (line \ref{alg:louvainlm--best-community-begin}), calculate the delta-modularity of moving $i$ to $c*$ (line \ref{alg:louvainlm--best-community-end}), and update the community membership  (lines \ref{alg:louvainlm--perform-move-begin}-\ref{alg:louvainlm--perform-move-end}) of $i$, and mark its neighbors as unprocessed (line \ref{alg:louvainlm--remark}) if a better community was found. In line \ref{alg:louvainlm--locally-converged}, we check if the local-moving phase has converged (locally). If so, we break out of the loop (or if $MAX\_ITERATIONS$ is reached). At the end, in line \ref{alg:louvainlm--return}, we return the number of iterations performed $l_i$.

\subsubsection{Aggregation phase of GVE-Louvain}

Finally, the psuedocode for the aggregation phase is shown in Algorithm \ref{alg:louvainag}, which aggregates communities into super-vertices. Here, the \texttt{louvainAggrega} \texttt{te()} function takes the current graph $G'$ and the community membership $C'$ as input, and returns the super-vertex graph $G''$.

In lines \ref{alg:louvainag--coff-begin}-\ref{alg:louvainag--coff-end}, the offsets array for the community vertices CSR $G'_{C'}.offsets$ is obtained by first counting the number of vertices belonging to each community using \texttt{countCommunity} \texttt{Vertices()}, and then performing exclusive scan on the array. In lines \ref{alg:louvainag--comv-begin}-\ref{alg:louvainag--comv-end}, we iterate over all vertices in parallel and atomically populate vertices belonging to each community into the community graph CSR $G'_{C'}$. Next, we obtain the offsets array for the super-vertex graph CSR by over-estimating the degree of each super-vertex, i.e., by obtaining the total degree of each community with \texttt{communityTotalDegree()}, and then performing exclusive scan on the array (lines \ref{alg:louvainag--yoff-begin}-\ref{alg:louvainag--yoff-end}). This causes the super-vertex graph CSR to be holey, i.e., with gaps in between the edges and weights array of each super-vertex in the CSR. Then, in lines \ref{alg:louvainag--y-begin}-\ref{alg:louvainag--y-end}, we iterate over all communities $c \in [0, |\Gamma|)$ in parallel, and add all communities $d$ (with associated edge weight $w$) linked to each vertex $i$ belonging to community $c$ (with \texttt{scanCommunities()} defined in Algorithm \ref{alg:louvainlm}) to the per-thread hashtable $H_t$. Once $H_t$ is populated with all communities (and associated weights) linked to community $c$, we add them as edges to super-vertex $c$ into the super-vertex graph $G''$ atomically. At the end, in line \ref{alg:louvainag--return}, we return the super-vertex graph $G''$.

\section{Evaluation}
\label{sec:evaluation}
\subsection{Experimental Setup}
\label{sec:setup}

\subsubsection{System used}

We use a server that has two $16$-core x86-based Intel Xeon Gold 6226R processors running at $2.90$ GHz. Each core has an L1 cache of $1$ MB, an L2 cache of $16$ MB, and a shared L3 cache of $22$ MB. The machine has $93.4$ GB of system memory and runs on CentOS Stream 8. For our GPU experiments, we employ a system featuring an NVIDIA A100 GPU (108 SMs, 64 CUDA cores per SM, 80 GB global memory, 1935 GB/s bandwidth, 164 KB shared memory per SM) paired with an AMD EPYC-7742 processor (64 cores, 2.25 GHz). This server is equipped with 512 GB of DDR4 RAM and operates on Ubuntu 20.04.

\subsubsection{Configuration}

We employ 32-bit integers for vertex IDs and 32-bit floats for edge weights, but use 64-bit floats for both computations and hashtable values. Our implementation leverages $64$ threads to align with the number of cores on the system, unless otherwise specified. For compilation, we use GCC 8.5 and OpenMP 4.5 on the CPU system, while on the GPU system, we use GCC 9.4, OpenMP 5.0, and CUDA 11.4.

\subsubsection{Dataset}

The graphs utilized in our experiments are detailed in Table \ref{tab:dataset}. These graphs are sourced from the SuiteSparse Matrix Collection \cite{suite19}. The graphs have vertex counts ranging from $3.07$ to $214$ million and edge counts ranging from $25.4$ million to $3.80$ billion. We ensure that the edges are undirected and weighted, with a default weight of $1$.

\begin{table}[hbtp]
  \centering
  \caption{List of $13$ graphs obtained SuiteSparse Matrix Collection \cite{suite19} (directed graphs are marked with $*$). Here, $|V|$ is the number of vertices, $|E|$ is the number of edges (after adding reverse edges), $D_{avg}$ is the average degree, and $|\Gamma|$ is the number of communities obtained using GVE-Louvain.}
  \label{tab:dataset}
  \begin{tabular}{|c||c|c|c|c|}
    \toprule
    \textbf{Graph} &
    \textbf{\textbf{$|V|$}} &
    \textbf{\textbf{$|E|$}} &
    \textbf{\textbf{$D_{avg}$}} &
    \textbf{\textbf{$|\Gamma|$}} \\
    \midrule
    \multicolumn{5}{|c|}{\textbf{Web Graphs (LAW)}} \\ \hline
    indochina-2004$^*$ & 7.41M & 341M & 41.0 & 4.24K \\ \hline  
    uk-2002$^*$ & 18.5M & 567M & 16.1 & 42.8K \\ \hline  
    arabic-2005$^*$ & 22.7M & 1.21B & 28.2 & 3.66K \\ \hline  
    uk-2005$^*$ & 39.5M & 1.73B & 23.7 & 20.8K \\ \hline  
    webbase-2001$^*$ & 118M & 1.89B & 8.6 & 2.76M \\ \hline  
    it-2004$^*$ & 41.3M & 2.19B & 27.9 & 5.28K \\ \hline  
    sk-2005$^*$ & 50.6M & 3.80B & 38.5 & 3.47K \\ \hline  
    \multicolumn{5}{|c|}{\textbf{Social Networks (SNAP)}} \\ \hline
    com-LiveJournal & 4.00M & 69.4M & 17.4 & 2.54K \\ \hline  
    com-Orkut & 3.07M & 234M & 76.2 & 29 \\ \hline  
    \multicolumn{5}{|c|}{\textbf{Road Networks (DIMACS10)}} \\ \hline
    asia\_osm & 12.0M & 25.4M & 2.1 & 2.38K \\ \hline  
    europe\_osm & 50.9M & 108M & 2.1 & 3.05K \\ \hline  
    \multicolumn{5}{|c|}{\textbf{Protein k-mer Graphs (GenBank)}} \\ \hline
    kmer\_A2a & 171M & 361M & 2.1 & 21.2K \\ \hline  
    kmer\_V1r & 214M & 465M & 2.2 & 6.17K \\ \hline  
  \bottomrule
  \end{tabular}
\end{table}

\begin{figure*}[hbtp]
  \centering
  \subfigure[Runtime in seconds (logarithmic scale) with \textit{Vite (Louvain)}, \textit{Grappolo (Louvain)}, \textit{NetworKit Louvain}, \textit{cuGraph Louvain}, and \textit{GVE-Louvain}]{
    \label{fig:louvain-compare--runtime}
    \includegraphics[width=0.98\linewidth]{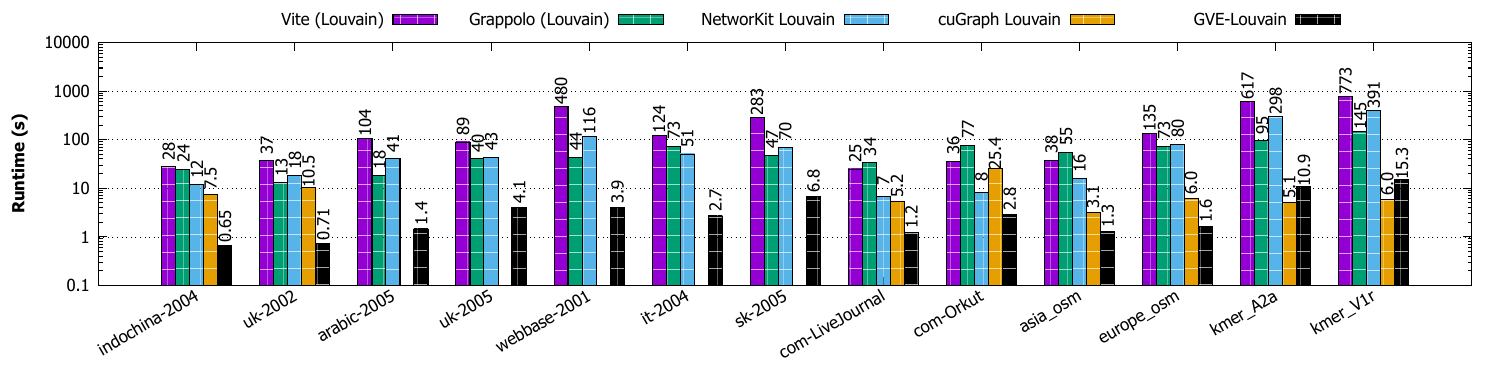}
  } \\[-0ex]
  \subfigure[Speedup of \textit{GVE-Louvain} with respect to \textit{Vite (Louvain)}, \textit{Grappolo (Louvain)}, \textit{NetworKit Louvain}, and \textit{cuGraph Louvain}.]{
    \label{fig:louvain-compare--speedup}
    \includegraphics[width=0.98\linewidth]{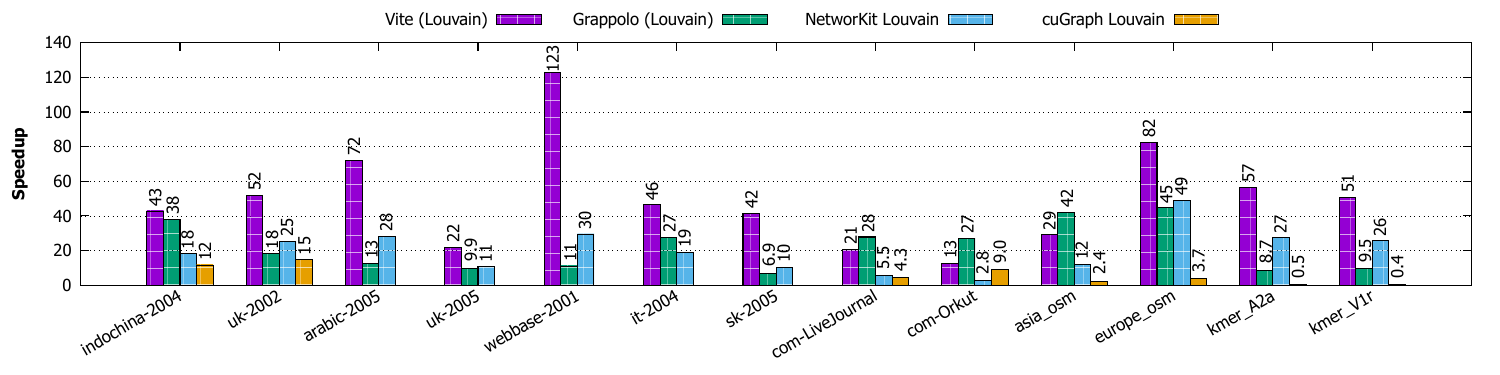}
  } \\[-0ex]
  \subfigure[Modularity of communities obtained with \textit{Vite (Louvain)}, \textit{Grappolo (Louvain)}, \textit{NetworKit Louvain}, \textit{cuGraph Louvain}, and \textit{GVE-Louvain}.]{
    \label{fig:louvain-compare--modularity}
    \includegraphics[width=0.98\linewidth]{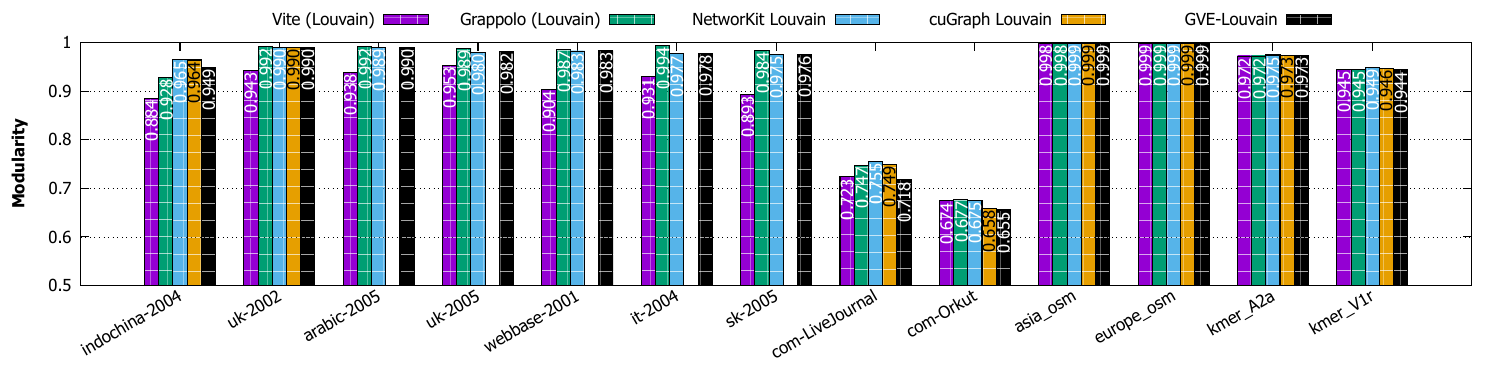}
  } \\[-2ex]
  \caption{Runtime in seconds (logarithmic scale), speedup, and modularity of communities obtained with \textit{Vite (Louvain)}, \textit{Grappolo (Louvain)}, \textit{NetworKit Louvain}, \textit{cuGraph Louvain}, and \textit{GVE-Louvain} for each graph in the dataset.}
  \label{fig:louvain-compare}
\end{figure*}

\subsection{Comparing Performance of GVE-Louvain}
\label{sec:comparison}

We now compare the performance of GVE-Louvain with Vite (Louvain) \cite{ghosh2018scalable}, Grappolo (Louvain) \cite{com-halappanavar17}, NetworKit Louvain \cite{staudt2016networkit}, and cuGraph Louvain \cite{kang2023cugraph}. For Vite, we convert the graph datasets to Vite's binary graph format, run it on a single node\ignore{(Vite supports distributed community detection)} with threshold cycling/scaling optimization, and measure the reported average total time. For Grappolo, we measure the run it on the same system, and measure the reported total time. For NetworKit Louvain, we use a Python script to invoke \texttt{PLM} (Parallel Louvain Method), and measure the total time reported with \texttt{getTiming()}. To test cuGraph's Louvain algorithm, we write a Python script that configures the Rapids Memory Manager (RMM) to use a pool allocator for fast memory allocations. We then execute \texttt{cugraph.louvain()} on the loaded graph. For each graph, we measure the runtime and the modularity of the obtained communities (as reported by each implementation), performing five runs to calculate an average. When using cuGraph, we discard the runtime of the first run to ensure that subsequent measurements accurately reflect RMM's pool usage without the overhead of initial CUDA memory allocation.

Figure \ref{fig:louvain-compare--runtime} shows the runtimes of Vite (Louvain), Grappolo (Louvain), NetworKit Louvain, cuGraph Louvain, and GVE-Louvain on each graph in the dataset. cuGraph's Louvain algorithm fails to run on the \textit{arabic-2005}, \textit{uk-2005}, \textit{webbase-2001}, \textit{it-2004}, and \textit{sk-2005} graphs because of out-of-memory issues. On the \textit{sk-2005} graph, GVE-Louvain finds communities in $6.8$ seconds, and thus achieve a processing rate of $560$ million edges/s. Figure \ref{fig:louvain-compare--speedup} shows the speedup of GVE-Louvain with respect to each implementation mentioned above. GVE-Louvain is on average $50\times$, $22\times$, $20\times$, and $5.8\times$ faster than Vite, Grappolo, NetworKit Louvain, and cuGraph Louvain respectively.  Figure \ref{fig:louvain-compare--modularity} shows the modularity of communities obtained with each implementation. GVE-Louvain on average obtains $3.1\%$ higher modularity than Vite (especially on web graphs), and $0.6\%$ lower modularity than Grappolo and NetworKit (especially on social networks with poor clustering), and $2.6\%$ higher modularity than cuGraph Louvain (primarily because cuGraph Louvain failed to run on graphs that are well-clusterable).

\begin{figure*}[hbtp]
  \centering
  \subfigure[Phase split]{
    \label{fig:louvain-splits--phase}
    \includegraphics[width=0.48\linewidth]{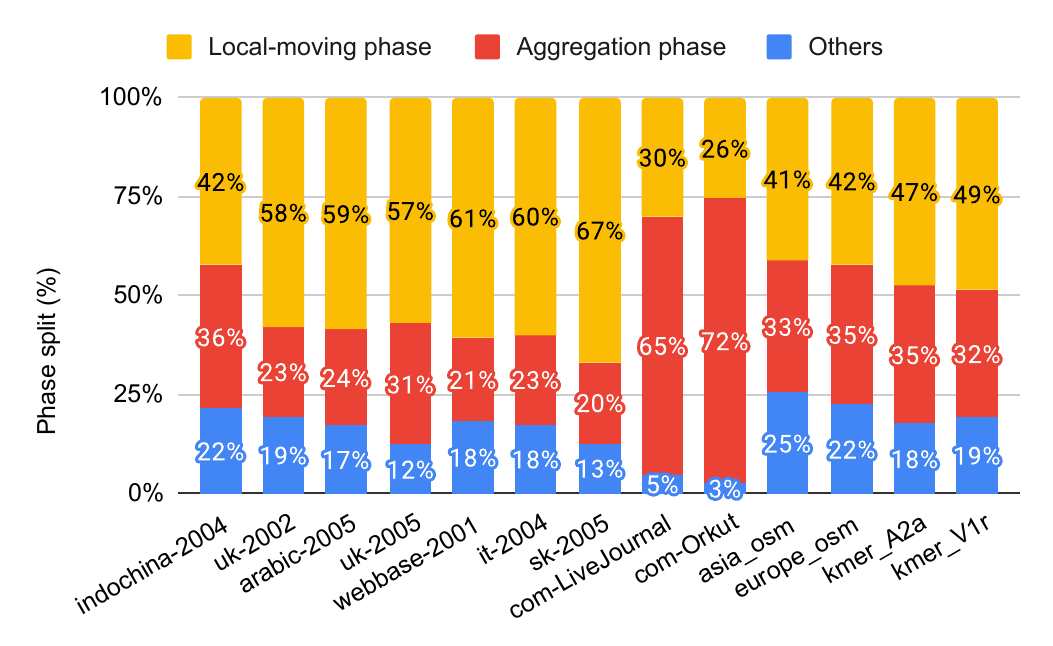}
  }
  \subfigure[Pass split]{
    \label{fig:louvain-splits--pass}
    \includegraphics[width=0.48\linewidth]{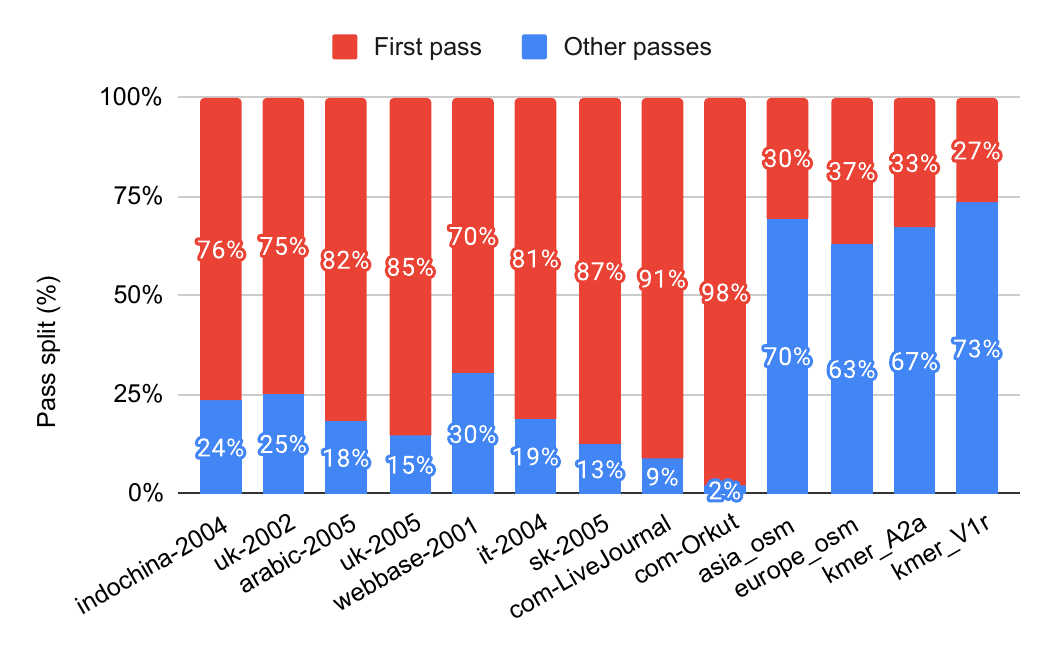}
  } \\[-2ex]
  \caption{Phase split of \textit{GVE-Louvain} shown on the left, and pass split shown on the right for each graph in the dataset.}
  \label{fig:louvain-splits}
\end{figure*}

\begin{figure}[hbtp]
  \centering
  \subfigure{
    \label{fig:louvain-hardness--all}
    \includegraphics[width=0.98\linewidth]{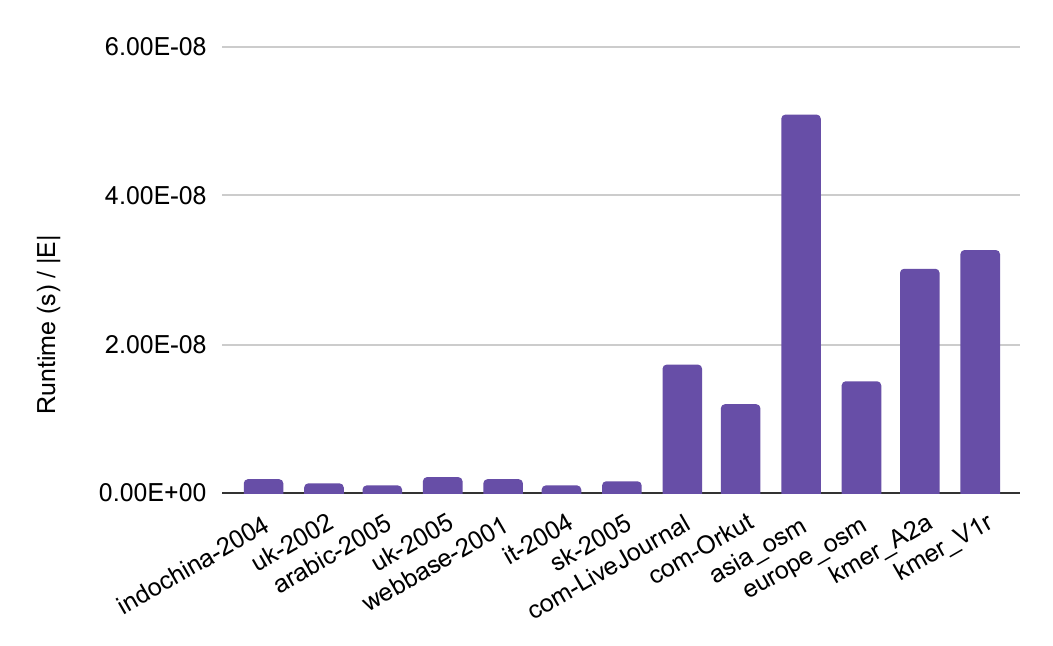}
  } \\[-2ex]
  \caption{Runtime $/ |E|$ factor with \textit{GVE-Louvain} for each graph in the dataset.}
  \label{fig:louvain-hardness}
\end{figure}

\begin{figure}[hbtp]
  \centering
  \includegraphics[width=0.98\linewidth]{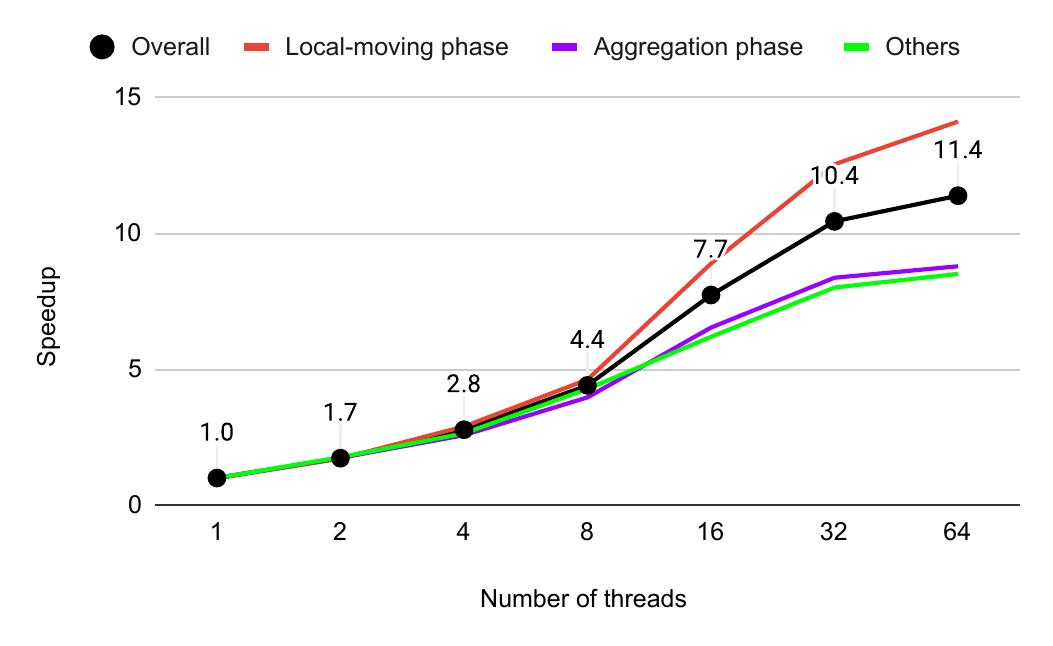} \\[-2ex]
  \caption{Overall speedup of \textit{GVE-Louvain}, and its various phases (local-moving, aggregation, others), with increasing number of threads (in multiples of 2).}
  \label{fig:louvain-ss}
\end{figure}

\subsection{Analyzing Performance of GVE-Louvain}

The phase-wise and pass-wise split of GVE-Louvain is shown in Figures \ref{fig:louvain-splits--phase} and \ref{fig:louvain-splits--pass} respectively. Figure \ref{fig:louvain-splits--phase} indicates that GVE-Louvain spends most of the runtime in the local-moving phase on \textit{web graphs}, \textit{road networks}, and \textit{protein k-mer graphs}, while it devotes majority of the runtime in the aggregation phase on \textit{social networks}. The pass-wise split (Figure \ref{fig:louvain-splits--pass}) indicates that the first pass dominates runtime on high-degree graphs (\textit{web graphs} and \textit{social networks}), while subsequent passes prevail in execution time on low-degree graphs (\textit{road networks} and \textit{protein k-mer graphs}).

On average, $49\%$ of GVE-Louvain's runtime is spent in the local-moving phase, $35\%$ is spent in the aggregation phase, and $16\%$ is spent in other steps (initialization, renumbering communities, looking up dendrogram, and resetting communities) of the algorithm. Further, $67\%$ of the runtime is spent in the first pass of the algorithm, which is the most expensive pass due to the size of the original graph (later passes work on super-vertex graphs) \cite{com-wickramaarachchi14}.

We also observe that graphs with lower average degree (\textit{road networks} and \textit{protein k-mer graphs}) and graphs with poor community structure (such as \verb|com-LiveJournal| and \verb|com-Orkut|) have a larger $\text{runtime}/|E|$ factor, as shown in Figure \ref{fig:louvain-hardness}.

\subsection{Strong Scaling of GVE-Louvain}

Finally, we measure the strong scaling performance of GVE-Louvain. To this end, we adjust the number of threads from $1$ to $64$ in multiples of $2$ for each input graph, and measure the overall time taken for finding communities with GVE-Louvain, as well as its phase splits (local-moving, aggregation, others), five times for averaging. The results are shown in Figure \ref{fig:louvain-ss}. With 32 threads, GVE-Louvain obtains an average speedup of $10.4\times$ compared to running with a single thread, i.e., its performance increases by $1.6\times$ for every doubling of threads. Scaling is limited due to the various sequential steps/phases in the algorithm. At 64 threads, GVE-Louvain is impacted by NUMA effects, and offers speedup of only $11.4\times$.

\section{Conclusion}
\label{sec:conclusion}
In conclusion, this report presents our parallel multicore implementation of the Louvain algorithm --- a high quality community detection method, which, as far as we are aware, stands as the most efficient implementation of the algorithm on multicore CPUs. Here, we considered $9$ different optimizations, which significantly improve the performance of the local-moving and the aggregation phases of the algorithm. On a server with dual 16-core Intel Xeon Gold 6226R processors, comparison with competitive open source implementations (Vite and Grappolo) and packages (NetworKit) show that our optimized implementation of Louvain, which we term as GVE-Louvain, is on average $50\times$, $22\times$, $20\times$, and $5.8\times$ faster than Vite, Grappolo, NetworKit Louvain, and cuGraph Louvain (run on an NVIDIA A100 GPU) respectively. In addition, GVE-Louvain on average obtains $3.2\%$ higher modularity than Vite\ignore{(especially on web graph)}, $0.6\%$ lower modularity than Grappolo and NetworKit Louvain\ignore{(especially on social networks with poor clustering)}, and $2.6\%$ higher modularity than cuGraph Louvain. On a web graph with $3.8$ billion edges, GVE-Louvain identifies communities in $6.8$ seconds, and thus achieves a processing rate of $560$ million edges/s. In addition, GVE-Louvain achieves a strong scaling factor of $1.6\times$ for every doubling of threads. Future work could focus of designing fast community detection algorithms that enable interactive updation of community memberships of vertices in large dynamic graphs.

\begin{acks}
I would like to thank Prof. Kishore Kothapalli,  Prof. Dip Sankar Banerjee, Balavarun Pedapudi, Souvik Karfa, Chuck Hastings, and Rick Ratzel for their support.
\end{acks}

\bibliographystyle{ACM-Reference-Format}
\bibliography{main}

\clearpage
\appendix
\section{Appendix}
\subsection{Indirect Comparison with State-of-the-art Louvain Implementations}
\label{sec:comparison-indirect}

We now conduct an indirect comparison of the performance of our multicore implementation of the Louvain algorithm with other similar state-of-the-art implementations, as listed in Table \ref{tab:compare}. Mohammadi et al. \cite{com-mohammadi20} introduce the Adaptive CUDA Louvain Method (ACLM), which employs GPU acceleration. For a computational setup featuring a $12$-core AMD Opteron 6344 CPU clocked at $2.60$ GHz alongside an NVIDIA Tesla K20Xm GPU, Mohammadi et al. report a runtime of $0.47$ seconds for ACLM when applied to the \textit{in-2004} graph with $13.6$ million edges (refer to Table 6 in \cite{com-mohammadi20}). Moreover, they document runtimes of $3.71$, $0.60$, and $0.95$ seconds for PLM \cite{staudt2015engineering}, APLM \cite{com-fazlali17}, and Rundemanen \cite{com-naim17} respectively (also presented in Table 6 of \cite{com-mohammadi20}). In contrast, leveraging our system equipped with two $16$-core Intel Xeon Gold 6226R CPUs operating at $2.90$ GHz (delivering up to $3.0\times$ higher performance), we process the \textit{indochina-2004} graph boasting $341$ million edges ($25.1\times$ larger) in merely $0.65$ seconds. Consequently, our Louvain implementation outperforms ACLM, PLM, APLM, and Rundemanen by approximately $6.0\times$, $48\times$, $7.7\times$, and $12.2\times$ respectively\ignore{ (all achieved without GPU)}.

Sattar and Arifuzzaman \cite{sattar2022scalable} propose their Distributed Parallel Louvain Algorithm with Load-balancing (DPLAL). On a graph containing $1M$ vertices, utilizing $40$ compute nodes of the Louisiana Optical Network Infrastructure (LONI) QB2 cluster, where each node is equipped with two $10$ core $2.80$ GHz Intel Xeon E5-2680v2 CPUs, they achieve community detection within approximately $80$ seconds (refer to Figure 4 in their paper \cite{sattar2022scalable}). In contrast, employing our system\ignore{featuring two $16$ core Intel Xeon Gold 6226R CPUs operating at $2.90$ GHz} (which is $6.0\times$ slower, assuming DPLAL utilizes each node with only $25\%$ efficiency), we process the \textit{indochina-2004} graph with $7.41M$ vertices ($7.41\times$ larger) in only $0.65$ seconds. Consequently, our Louvain implementation is roughly $5472\times$ faster\ignore{ than DPLAL}\ignore{ exhibits a speedup of roughly $5472\times$ compared to DPLAL}.

Qie et al. \cite{qie2022isolate} introduce a graph partitioning algorithm designed to minimize inter-partition communication delay and avoid community swaps. Their experiments on an $8$ core Intel i9 9900K CPU running at $3.60$ GHz, using the \textit{com-LiveJournal} graph, reveal a runtime of approximately $1050$ seconds (as depicted in Figure 3 of their paper \cite{qie2022isolate}). In contrast, on our system\ignore{ leveraging our system equipped with two $16$ core Intel Xeon Gold 6226R CPUs running at $2.90$ GHz} (roughly $3.2\times$ faster), we process the same graph in merely $1.2$ seconds. Consequently, our Louvain implementation achieves a speedup of about $273\times$.

Bhowmick et al. \cite{com-bhowmik19} present HyDetect, a community detection algorithm tailored for hybrid CPU-GPU systems, employing a divide-and-conquer strategy. Their experimentation on a system comprising a six-core Intel Xeon E5-2620 CPU operating at $2.00$ GHz paired with an NVIDIA Kepler K40M GPU yields a runtime of $1123$ seconds when applied to the \textit{it-2004} graph (as documented in Table 5 of their paper \cite{com-bhowmik19}). In contrast, leveraging our system equipped with two $16$ core Intel Xeon Gold 6226R CPUs running at $2.90$ GHz (approximately $7.7\times$ faster), we process the same graph in $2.7$ seconds. Consequently, our Louvain implementation achieves a speedup of around $54\times$ compared to HyDetect, without using a GPU.

Bhowmick et al. \cite{com-bhowmick22} later introduce a multi-node multi-GPU Louvain community detection algorithm. Their experimentation conducted on a CrayXC40 system, where each node comprises one $12$ core $2.40$ GHz Intel Xeon Ivybridge E5-2695 v2 CPU and one NVIDIA Tesla K40 GPU, demonstrates the capability to identify communities within approximately $32$ seconds on the \textit{uk-2005} graph utilizing $8$ compute nodes (refer to Figure 8 in their paper \cite{com-bhowmick22}). Furthermore, they observe a speedup of about $1.7\times$ compared to Cheong et al. \cite{com-cheong13} on the \textit{uk-2005} graph (as depicted in Figure 11 of their paper), and a speedup of approximately $1.2\times$ relative to Ghost et al. \cite{com-ghosh18} on the \textit{sk-2005} graph (as shown in Figure 10 of their paper). In contrast, on our system featuring two $16$ core Intel Xeon Gold 6226R CPUs operating at $2.90$ GHz (up to $1.6\times$ faster, assuming Bhowmick et al. utilize each compute node with only $25\%$ efficiency), we process the \textit{uk-2005} graph in a mere $4.1$ seconds. Thus, our Louvain implementation achieves at least $4.9\times$, $8.3\times$, and $5.9\times$ speedup compared to Multi-node HyDetect \cite{com-bhowmick22}, Cheong et al. \cite{com-cheong13}, and Ghost et al. \cite{com-ghosh18}, respectively, without\ignore{requiring} the use of a\ignore{single} GPU.

Chou and Ghosh \cite{chou2022batched} introduce Nido, a batched clustering method designed for GPUs. Their experimentation conducted on a system equipped with two $128$ core AMD EPYC 7742 CPUs running at $2.25$ GHz and eight NVIDIA A100 GPUs demonstrates a (geometric) mean speedup of $2.4\times$ compared to Grappolo \cite{com-halappanavar17}, utilizing $128$ threads (as shown in Table 3 of their paper \cite{chou2022batched}).\ignore{Additionally, they find that cuGraph \cite{hricik2020using} offers a mean speedup of $7.7\times$ relative to Grappolo (also depicted in Table 3 of their paper).} In contrast, our Louvain implementation achieves a mean speedup of $22\times$ compared to Grappolo. Consequently, it surpasses Chou and Ghosh by $9.2\times$\ignore{and cuGraph by $2.9\times$, again without employing a single GPU}.

Gawande et al. \cite{com-gawande22} introduce cuVite, an ongoing project focusing on distributed Louvain for heterogeneous systems. They evaluate cuVite, cuGraph, and Rundemanen on a single-GPU of NVIDIA DGX-2, i.e., NVIDIA Tesla V100, and with two $24$ core Intel Xeon Platinum 8168 CPUs running at $2.7$ GHz. Further, they evaluate Grappolo on a $224$ core system with eight $28$ core Intel Xeon Platinum 8276M CPUs running at $2.20$ GHz. They observe that, on average,\ignore{cuGraph,} Rundemanen and cuVite exhibit speedups of\ignore{$3.3\times$,} $3.0\times$ and $3.28\times$ respectively compared to Grappolo. \ignore{Further analysis on the \textit{webbase-2001} graph indicates that cuVite (32 nodes) achieves a speedup of $1.3\times$ (including remapping) or $2.0\times$ faster (excluding remapping) relative to Grappolo (128 threads), while Vite (32 nodes) demonstrates a speedup of 2.3 times. Notably, cuVite does not significantly outperform Vite, and in some cases, it exhibits slower performance.} Consequently, our Louvain implementation achieves a speedup of $6.7\times$ compared to\ignore{both cuGraph and} cuVite, and a speedup of $7.3\times$ relative to Rundemanen\ignore{, all achieved without the utilization of a GPU}.

\end{document}